\documentclass[twocolumn,aps,showpacs,prb,tightenlines,amsmath,amssymb]{revtex4}

\usepackage{graphicx}
\usepackage{amssymb} 
\usepackage{amsmath}
\usepackage{colordvi}
\newcommand{\bgreek}[1]{\mbox{\boldmath$#1$\unboldmath}}
\begin{document} 

\title{Hole spin relaxation in [001] strained asymmetric Si/SiGe and
Ge/SiGe quantum wells}
\author{P. Zhang}
\affiliation{Hefei National Laboratory for Physical Sciences at
  Microscale and Department of Physics, 
University of Science and Technology of China, Hefei,
  Anhui, 230026, China}
\author{M. W. Wu}
\thanks{Author to whom correspondence should be addressed}
\email{mwwu@ustc.edu.cn.}
\affiliation{Hefei National Laboratory for Physical Sciences at
  Microscale and Department of Physics, 
University of Science and Technology of China, Hefei,
  Anhui, 230026, China}

\date{\today}
\begin{abstract} 
Hole spin relaxation in [001] strained asymmetric
Si/Si$_{0.7}$Ge$_{0.3}$ (Ge/Si$_{0.3}$Ge$_{0.7}$) quantum wells is investigated in the
situation with only the lowest hole subband
being relevant. The effective
Hamiltonian of the lowest hole subband is obtained by the subband L\"owdin
perturbation method in the framework of the six-band 
Luttinger ${\bf k}\cdot{\bf p}$
model, with sufficient basis functions included. The lowest hole subband in Si/SiGe quantum wells is light-hole
like with the Rashba spin-orbit coupling term depending on momentum both
linearly and cubically, while that in Ge/SiGe quantum wells
is a heavy hole state with the Rashba spin-orbit coupling term
depending on momentum only cubically. The hole spin relaxation is investigated
by means of the fully microscopic kinetic spin Bloch equation
approach, with all the relevant scatterings considered. It is found that
 the hole-phonon scattering is very weak, which makes the hole-hole Coulomb
scattering become very important. The hole system in Si/SiGe quantum wells is
  generally in the strong scattering limit, while that in Ge/SiGe
  quantum wells can be in either the
  strong or the weak scattering limit. The Coulomb scattering leads to a
  peak in both the temperature and hole density dependences of spin
  relaxation time in Si/SiGe quantum wells, located around the crossover between
  the degenerate and nondegenerate regimes. Nevertheless, the Coulomb
  scattering leads to not only a peak but also a valley in the
  temperature dependence of spin relaxation time in Ge/SiGe quantum wells. The
  valley is actually due to the crossover from the weak to strong scattering
  limit. The hole-impurity scattering influences the spin relaxation
  effectively. In the strong (weak) scattering limit, the spin
  relaxation time increases (decreases) with increasing impurity density. The
  spin relaxation time is found to be on the order of 1$\sim$100~ps
  (0.1$\sim$10~ps) in Si/SiGe (Ge/SiGe) quantum wells, for the temperatures, carrier/impurity
  densities and gate voltages of our consideration.
 
\end{abstract}
\maketitle

\section{Introduction}
In recent years, great efforts have been devoted to the design/realization of
spintronic devices, which employ the spin
degree of freedom in traditional electronics for the sake of higher power
efficiency, higher speed, and also greater functionality.\cite{aws,zutic,dy}
Among different kinds of hosts for 
such devices, Si appears to be a particularly promising one and attracts much attention, partly due to the high
possibility of eliminating hyperfine couplings by isotopic
purification and the well developed microfabrication
technology.\cite{schaffler} In fact, the electron spin relaxation, which is necessary to be
understood for the device design, has been widely
investigated in Si materials during the last decade. The study on
relaxation of electron spin qubit in Si quantum dot suggests that
the relaxation rate can be strongly decreased
by adding strain.\cite{tahan0} The electron spin
relaxation in asymmetric $n$-type Si/SiGe quantum wells (QWs) has been
investigated both theoretically\cite{tahan} and
experimentally.\cite{wil,jantsch} It is 
shown that the electron spin relaxation time can be quite long (on the order of
$10^{-7}\sim10^{-5}$~s)\cite{tahan,wil,jantsch} due to the weak Rashba
spin-orbit coupling\cite{rashba} (typically
about three orders of magnitude smaller than that in QW structures
based on III-V semiconductors\cite{wil}). The electron spin
transport/diffusion in bulk Si with a magnetic
field perpendicular to both the directions of spin polarization and
spin transport/diffusion has also been studied recently.\cite{appelbaum}
 It is revealed
that even in the absence of the traditional D'yakonov-Perel' (DP)
relaxation mechanism,\cite{dp} there is an obvious spin relaxation along spin
transport/diffusion,  as predicted several years
ago from a general QW model without any DP relaxation
mechanism but with a magnetic field in the Voigt
configuration.\cite{weng2002} 
That is also the case in
the symmetric Si/SiGe QWs.\cite{zhang}

Although a broad interest has been taken in the electron spin
relaxation in Si, to our knowledge, the hole spin relaxation
 has been rarely investigated so far. Glavin and Kim have
calculated the spin relaxation of two-dimensional holes in strained
asymmetric Si/SiGe (Ge/SiGe) QWs four years ago,\cite{glavin} and obtained a
spin relaxation time of several tens of picoseconds (several
sub-picoseconds) in Si/SiGe (Ge/SiGe) QWs with large gate voltage
(which induces an electric field at $50\sim500$~kV/cm) 
at room temperature. However, the results were obtained
by means of the single-particle
approximation,\cite{sp} therefore the effect of 
the carrier-carrier Coulomb scattering on spin relaxation, 
which has been revealed to be important in spin
relaxation,\cite{wu-rev,glazov,zhou,ruan,leyland,brand,glazov1} was not
included. Besides, the nondegenerate perturbation method with only the
lowest unperturbed subband of each hole state considered as basis
function is utilized 
to calculate the subband energy spectrum and envelope functions in
Ref.~\onlinecite{glavin}. However, as shown later in this
paper, only considering the lowest unperturbed subband is inadequate
in converging the calculation,
but when more unperturbed  subbands are included 
as basis functions, the nondegenerate perturbation method fails.
This work is to perform a detailed investigation on hole spin
relaxation in asymmetric Si/SiGe and Ge/SiGe QWs by means of the fully microscopic kinetic
spin Bloch equation (KSBE) approach,\cite{wu-rev} with all the
relevant scatterings included. Meanwhile, we apply the exact 
diagonalization method to obtain the
 energy spectrum and envelope functions, with sufficient unperturbed
subbands included. In the KSBE approach, the momentum-dependent spin precessions
give rise to the inhomogeneous broadening, with which
any scattering (including the Coulomb
scattering) leads to an irreversible spin relaxation.\cite{wu-rev}
This approach has been successfully applied 
to study spin dynamics in  quantum 
wire,\cite{cheng,lv} QW\cite{yzhou,zhang,weng,zhou,weng2002,lv1} 
and bulk\cite{jiang} semiconductor
structures. The current work reveals that the Coulomb scattering plays
a much more important role in hole spin relaxation in Si/SiGe
(Ge/SiGe) QWs. It leads to a peak in both the temperature and density
dependences of spin relaxation time in Si/SiGe QWs, where holes are 
generally in the strong scattering limit. Nonetheless, it leads to 
not only a peak but also a valley in the temperature
dependence of spin relaxation time in Ge/SiGe QWs, where with the
change of temperature the holes in Ge/SiGe QWs can be 
in the either strong or weak scattering limit. 
Besides, the spin relaxation time can be
effectively influenced by the hole-impurity scattering, which tends to
weaken the effect of the Coulomb scattering mentioned above with the
increase of impurity density.

This paper is organized as follows. 
In Sec.~II the effective Hamiltonian of the lowest hole subband (we focus 
on the situations with only the lowest subband being 
relevant) in asymmetric Si/SiGe 
(Ge/SiGe) QWs is derived. In Sec.~III the KSBEs are
constructed and the hole spin relaxation in Si/SiGe (Ge/SiGe)
QWs is investigated. Finally, we conclude in Sec.~IV.

\section{Effective Hamiltonian}
\begin{figure}[t]
    {\includegraphics[width=8cm]{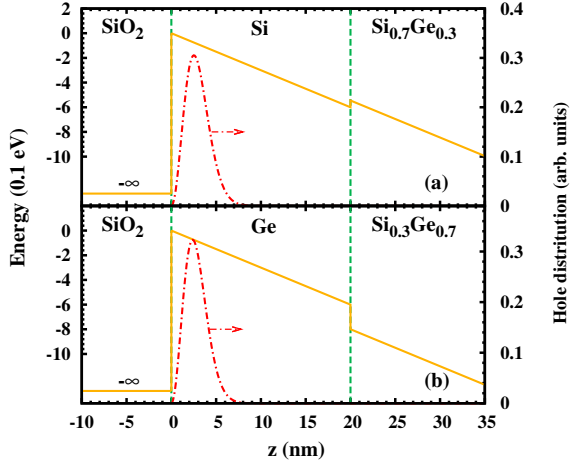}}
 \caption{(Color online) Schematics of the
 SiO$_2$/Si/Si$_{0.7}$Ge$_{0.3}$ QW structure (a) and
  SiO$_2$/Ge/Si$_{0.3}$Ge$_{0.7}$ QW structure (b). 
Two vertical dashed
 lines in each figure represent the two interfaces. The solid curves
represent the confining potential $V(z)$ with electric field $E=300$~kV/cm. 
The valence band discontinuities
at the Si/Si$_{0.7}$Ge$_{0.3}$ and Ge/Si$_{0.3}$Ge$_{0.7}$ interfaces 
are neglected in the triangular potential approximation. The chain 
curves with their   scale on the right hand side of the
   frame are $|\Psi_{1l0}(z)|^2$ (a) and
     $|\Psi_{1h0}(z)|^2$ (b), respectively,
 representing the lowest LH and HH distributions in Si/Si$_{0.7}$Ge$_{0.3}$
and Ge/Si$_{0.3}$Ge$_{0.7}$ QWs along the $z$-axis.}
  \label{figzw1}
\end{figure}

We start our investigation from the $p$-type
SiO$_2$/Si/Si$_{0.7}$Ge$_{0.3}$ (SiO$_2$/Ge/Si$_{0.3}$Ge$_{0.7}$)
QWs. The SiO$_2$/Si/Si$_{0.7}$Ge$_{0.3}$ and
SiO$_2$/Ge/Si$_{0.3}$Ge$_{0.7}$ QW structures are illustrated
in Fig.~\ref{figzw1}. The Si (Ge) layer is
[001]$||z$ grown with a wide width ($\ge 10$~nm). The SiO$_2$ layer is assumed to
be an infinite potential barrier. With the valence band discontinuity
at the Si/Si$_{0.7}$Ge$_{0.3}$ (Ge/Si$_{0.3}$Ge$_{0.7}$)
interface [$\approx 55$~meV ($\approx 200$~meV)]\cite{schaffler} 
ignored due to the large gate voltage
(inducing an electric field $\ge 50$~kV/cm) and
wide well width,\cite{glavin} 
the triangular potential approximation
is adopted\cite{glavin,fischetti,low} and the well width then becomes
irrelevant.

Based on the theory of Luttinger-Kohn\cite{luttinger1,luttinger2} and
Bir-Pikus,\cite{bir} the valence-band structure of the strained QWs can
be described by the 6$\times$6 effective-mass Hamiltonian,\cite{glavin,note1}
\begin{equation}
 H=H_L^{(0)}+H_L^{(||)}+H_{\epsilon}+V(z)I_6.
\label{eq1}
\end{equation}
Here $H_L\equiv H_L^{(0)}+H_L^{(||)}$ is the Luttinger
Hamiltonian\cite{luttinger1,luttinger2,bir} 
with $H_L^{(0)}$ corresponding to the part with
$k_{x,y}=0$. $H_\epsilon$ is the contribution due to the biaxial
strain.\cite{luttinger1,luttinger2,bir,chao} $V(z)$ is the
confining potential and $I_6$ is the
6$\times$6 unit matrix. The $z$-components of the
subband envelope functions (the $x$- and
$y$-components are plane waves) obtained by solving the eigen-equation of
$H_0=H_L^{(0)}+H_\epsilon+V(z)I_6$ are labeled as
\begin{eqnarray}\nonumber
\Psi_{1ln}(z)=\left(\begin{array}{c}0\\\chi_n^{(l1)}(z)\\0\\0\\i\chi_n^{(l2)}(z)\\0\end{array}
  \right),\Psi_{2ln}(z)=\left(\begin{array}{c}0\\0\\\chi_n^{(l1)}(z)\\0\\0\\i\chi_n^{(l2)}(z) \end{array}\right),\\
\Psi_{1hn}(z)=\left(\begin{array}{c}\chi_n^{(h)}(z)\\0\\0\\0\\0\\0\end{array}\right),\Psi_{2hn}(z)=\left(\begin{array}{c}0\\0\\0\\\chi_n^{(h)}(z)\\0\\0\end{array}\right),\label{eq2}\\\nonumber
\Psi_{1sn}(z)=\left(\begin{array}{c}0\\\chi_n^{(s1)}(z)\\0\\0\\i\chi_n^{(s2)}(z)\\0
  \end{array}\right),\Psi_{2sn}(z)=\left(\begin{array}{c}0\\0\\\chi_n^{(s1)}(z)\\0\\0\\i\chi_n^{(s2)}(z)\end{array}\right).
\end{eqnarray}
Here $l$, $h$ and $s$ represent the light
hole (LH), heavy hole (HH), and split-off (SO) hole states,
respectively, and $n$ is the subband number. The solution of the
envelope functions is stated in Appendix~A.

\begin{table}
\caption{Material parameters of Si and Ge. The mass density $d$, 
deformation potentials $D_{ac}$ and
  $\Delta_{op}$, optical phonon energy $\hbar\omega_{op}$ and 
  sound velocity $v_s$ are taken from 
Ref.~\onlinecite{bufler}. The Luttinger parameters
  $\gamma_1$, $\gamma_2$, $\gamma_3$ are from Ref.~\onlinecite{low}.}

\begin{tabular}{cccccc}
\hline\hline
          &   $d$      &  $D_{ac}$  &  $\Delta_{op}$    &  $\hbar\omega_{op}$  &  $v_s$\\
Material  & (g/cm$^3$)  &     (eV)   &  (10$^8$ eV/cm)  &  (eV)  &(10$^5$ cm/s)\\
\hline
Si  &    2.33          &  5.03            &   8.7  &      0.063  & 9.0\\      
Ge  &    5.32          &  3.5             &   7.0  &      0.037  & 5.4\\
\hline\hline
         &$\gamma_1$ & $\gamma_2$ & $\gamma_3$ \\
\hline
Si  &    4.285    &     0.339    &   1.446 \\
Ge  &    13.38    &     4.24     &   5.69  \\
\hline\hline
\end{tabular}
\label{table1}
\end{table}

The L\"owdin partitioning\cite{winkler} is performed upto 
second order in $H_L^{(||)}$ on
the basis constructed by $\Psi_{\lambda\alpha n}$ ($\lambda$=1, 2; 
$\alpha$=$h$, $l$, $s$) to obtain the effective Hamiltonian 
   of the lowest hole subband.\cite{note2} Due to the biaxial
strain,\cite{luttinger1,luttinger2,bir,chao} the lowest subband in Si/SiGe
QWs is a LH-like state (LH0), which is an admixture of LH and SO hole
states, while that in Ge/SiGe QWs is a pure HH state (HH0). The
effective Hamiltonian of the lowest hole subband in Si/SiGe (Ge/SiGe) 
QWs can be written as\cite{glavin}
\begin{equation}
  H^{(l,h)}_{\rm eff}=-\frac{\hbar^2{\bf k}^2}{2m^{(l,h)}}-\frac{\hbar}{2}{\bgreek\sigma}\cdot{\bf
    \Omega}^{(l,h)}(k_x,k_y),
\label{eqa3}
\end{equation}
where ${\bf k}$ is the in-plane momentum, $m^{(l)}$
[$m^{(h)}$] is the in-plane effective mass of the lowest light (heavy)
hole subband in Si/SiGe (Ge/SiGe) QWs,
$\bgreek{\sigma}$ are the Pauli matrices, and ${\bf
  \Omega}^{(l)}$ [${\bf \Omega}^{(h)}$] is the Rashba term of
 the LH0 (HH0) subband in Si/SiGe (Ge/SiGe) QWs. ${\bf\Omega}^{(l)}$ has both the
linear and cubic dependences on momentum, whereas ${\bf\Omega}^{(h)}$
has only the cubic dependence. For the LH0 subband in Si/SiGe QWs,
\begin{eqnarray}
m^{(l)}&=&m_0[A-B(\lambda^{(l1l1)}_{00}/2-\sqrt{2}\lambda^{(l1l2)}_{00})]^{-1},\label{ml}\\
{\bf \Omega}^{(l)}&=&{\bf \Omega}^{(l)}_1+{\bf \Omega}^{(l)}_3,\\
\Omega_{1x,y}^{(l)}&=&\Xi k_{x,y},\\\nonumber
\Omega_{3x}^{(l)}&=&\Pi Bk_x(k_x^2+k_y^2)+\Theta [3Bk_x(k_x^2-k_y^2)\\
&&+2\sqrt{3(3B^2+C^2)}k_y^2k_x],\\\nonumber
\Omega_{3y}^{(l)}&=&\Pi Bk_y(k_x^2+k_y^2)+\Theta [3Bk_y(k_y^2-k_x^2)\\
&&+2\sqrt{3(3B^2+C^2)}k_x^2k_y],
\end{eqnarray}
with
\begin{eqnarray}
\Xi&=&\frac{\hbar}{m_0}\sqrt{6(3B^2+C^2)}\kappa_{00}^{(l1l2)},\label{xi}\\\nonumber
  \Pi&=&-\frac{\hbar^3}{2m_0^2}\sqrt{\frac{3(3B^2+C^2)}{2}}\sum_{\alpha=l,s}\sum_{n=0}^{\infty}(1-\delta_{l\alpha}\delta_{0n})\\\nonumber &&\times\frac{\kappa_{0n}^{(l1\alpha 2)}-\kappa_{0n}^{(l2\alpha 1)}}{E_0^{(l)}-E_n^{(\alpha)}}[\sqrt{2}(\lambda_{0n}^{(l1\alpha 2)}+\lambda_{0n}^{(l2\alpha 1)})\\ &&-\lambda_{0n}^{(l1\alpha 1)}],\label{pi}\\\nonumber
\Theta&=&-\frac{\hbar^3}{2m_0^2}\sqrt{\frac{3B^2+C^2}{3}}\sum_{n=0}^{\infty}\frac{\kappa_{0n}^{(l1h)}-\frac{1}{\sqrt{2}}\kappa_{0n}^{(l2h)}}{E_0^{(l)}-E_n^{(h)}}\\
&&\times(\sqrt{2}\lambda_{0n}^{(l2h)}+\lambda_{0n}^{(l1h)})\label{theta}.
\end{eqnarray}
For the HH0 subband in Ge/SiGe QWs,
\begin{eqnarray}
m^{(h)}&=&m_0(A+B/2)^{-1},\label{mh}\\
{\bf \Omega}^{(h)}&=&{\bf \Omega}^{(h)}_3,\\\nonumber
\Omega_{3x}^{(h)}&=&\Lambda [3Bk_x(k_x^2-k_y^2)\\ &&-2\sqrt{3(3B^2+C^2)}k_y^2k_x],\\\nonumber
\Omega_{3y}^{(h)}&=&\Lambda [3Bk_y(k_x^2-k_y^2)\\ &&+2\sqrt{3(3B^2+C^2)}k_x^2k_y],
\end{eqnarray}
with
\begin{eqnarray}\nonumber
\Lambda
&=&-\frac{\hbar^3}{2m_0^2}\sqrt{\frac{3B^2+C^2}{3}}\sum_{\alpha=l,s}\sum_{n=0}^{\infty}\frac{\frac{1}{\sqrt{2}}\kappa_{0n}^{(h\alpha
    2)}-\kappa_{0n}^{(h\alpha 1)}}{E_0^{(h)}-E_n^{(\alpha)}}\\
&&\times(\sqrt{2}\lambda_{0n}^{(h\alpha 2)}+\lambda_{0n}^{(h\alpha
  1)}).
\label{lambda}
\end{eqnarray}
Here $A$, $B$ and $C$ are the valence band
parameters, which relate to the Luttinger parameters
(Table~\ref{table1}) $\gamma_1$,
$\gamma_2$ and $\gamma_3$ through $A=\gamma_1$, $B=2\gamma_2$ and
$\sqrt{3B^2+C^2}=2\sqrt{3}\gamma_3$. $E_n^{(\alpha)}$ ($\alpha$=$h$,
$l$, $s$) are the  subband energy
levels. $\lambda_{nn^\prime}^{(\alpha\beta)}$
and $\kappa_{nn^\prime}^{(\alpha\beta)}$ are defined as
$\lambda_{nn^\prime}^{(\alpha\beta)}=\int_{-\infty}^{+\infty}dz\chi_n^{(\alpha)}(z)\chi_{n^\prime}^{(\beta)}(z)$
and $\kappa_{nn^\prime}^{(\alpha\beta)}=\int_{-\infty}^{+\infty}dz\chi_n^{(\alpha)}(z)\frac{d\chi_{n^\prime}^{(\beta)}(z)}{dz}$. It is noted that in Ref.~\onlinecite{glavin} the coefficients $\Pi$, $\Theta$ and
$\Lambda$ miss the pre-factor $-\frac{1}{2}$  and meanwhile
$\Pi$ misses the summation over $l$ (i.e., the contribution due to the higher 
LH subbands, which is in fact negligibly small). Unlike the work by Glavin and
Kim\cite{glavin} where the nondegenerate perturbation method with only
the lowest unperturbed subband of each hole state being accounted [refer to
 Eqs.~(3--9) in Ref.~\onlinecite{glavin}]
 is employed in obtaining the subband energy spectrum $E^{(\alpha)}_{n}$
 and the envelope
 functions $\Psi_{\lambda\alpha n}$, we apply the exact diagonalization method
 with sufficient unperturbed subbands [2 (20) for each hole state in Si/SiGe (Ge/
SiGe) QWs] for the sake of convergence. In fact, the calculation
with only the lowest unperturbed subband of each hole state is {\em inadequate} for the
convergence. Moreover, the calculation with more unperturbed subbands included may cause
divergence as long as the nondegenerate perturbation method is
utilized. When performing the subband L\"owdin
  partition method\cite{winkler} to obtain the effective
 Hamiltonian of the lowest subband, we choose  sufficient envelope
functions $\Psi_{\lambda \alpha n}$  as basis functions. For Si/SiGe (Ge/SiGe)
 QWs, the number of envelope functions containing 
$\Psi_{\lambda l n}$ and $\Psi_{\lambda s n}$ is 4 (40) in
 total, and the number of envelope 
functions $\Psi_{\lambda h n}$ is 2 (20). The reason that more 
basis functions are needed for Ge/SiGe QWs 
comes from the fact that the couplings between the hole subbands are
stronger than those in the Si/SiGe QWs (the Luttinger parameters in Ge
are much larger than those in Si but under the strain the
energy differences between hole subbands are comparable in Si/SiGe and
Ge/SiGe QWs).

\begin{figure}[htb]
    {\includegraphics[width=9cm]{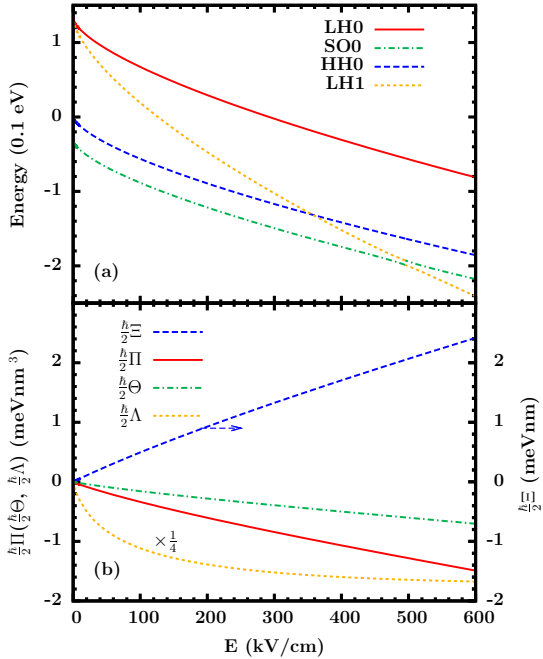}}
    \caption{(Color online) (a) The subband energy spectrum at $\Gamma$
      point against electric field in Si/SiGe QWs. Four subbands are
      shown: the first (second) light-hole subband LH0 (LH1), the
      first heavy-hole subband HH0 and the first split-off hole subband SO0.   
      (b) Spin-orbit coupling coefficients
      $\frac{\hbar}{2}\Xi$, $\frac{\hbar}{2}\Pi$ and 
      $\frac{\hbar}{2}\Theta$ for LH0 subband in Si/SiGe QWs and
      $\frac{\hbar}{2}\Lambda$ for HH0 subband in Ge/SiGe 
QWs against the electric field $E$. The
      scale of $\frac{\hbar}{2}\Xi$ is on the right hand side of the frame.}    
  \label{figzw2}
\end{figure} 

According to Eqs.~(\ref{ml}) and (\ref{mh}), the in-plane effective 
mass of LH0 subband
in Si/SiGe QWs $m^{(l)}$ is calculated to be about $0.27m_0$ in the whole electric 
field range under consideration, and that of the HH0 subband 
in Ge/SiGe QWs $m^{(h)}$ is $0.057m_0$.
We plot the energy levels of four subbands in Si/SiGe
QWs (the first and second LH subbands LH0 and
LH1, the first HH subband HH0, and the 
first SO subband SO0) at $\Gamma$ point in  Fig.~\ref{figzw2}(a)
 and the spin-orbit coupling
coefficients of the LH0 subband in Si/SiGe QWs
($\Xi$, $\Pi$ and $\Theta$) and the HH0 subband in Ge/SiGe QWs
($\Lambda$) in  Fig.~\ref{figzw2}(b). As shown in 
Fig.~\ref{figzw2}(a), a crossing between
the LH1 and HH0 subbands appears at about $E=360$~kV/cm and an 
anticrossing between the LH1 and SO0
subbands appears around $E=500$~kV/cm, respectively. It is noted that
notwithstanding the fact that our calculation goes to the 
infinitesimal electric field
regime in Fig.~\ref{figzw2}, only the results in the large electric
field regime (i.e., $E\ge 50$~kV/cm for QWs with well width
  $\sim 10$~nm) are valid (the spin relaxation investigated later is also in the
large electric field regime), as our model fails in the
small electric field regime due to the disregard of the
discontinuity in the Si/SiGe (Ge/SiGe) interface. In this work
the temperature dependence of band parameters\cite{zutic,Eld} is
not taken into account, due to both the
weak temperature dependence of band parameters and the
negligible contribution from the conduction band.\cite{note1}

\begin{figure}[thb]
 {\includegraphics[width=8.5cm]{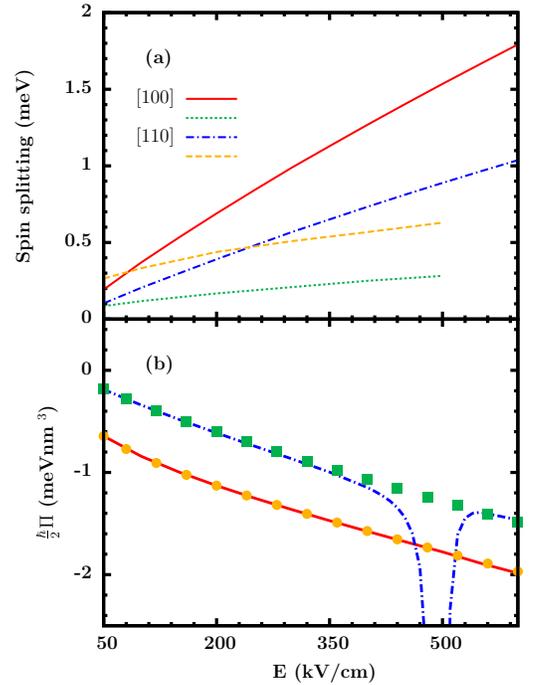}}
\caption{(Color online) (a) Spin splitting for holes with in-plane kinetic energy
 being $k_BT$ ($T=300$~K) along the [100] (the solid and dotted curves) 
and [110] (the chain and dashed curves) directions in Si/SiGe QWs. Solid
   and chain curves: results of our calculation; dotted and dashed curves:
   results from Ref.~\onlinecite{glavin}. (b) Spin-orbit coupling
      coefficient $\frac{\hbar}{2}\Pi$ calculated with the nondegenerate
perturbation and exact diagonalization methods
      respectively. The solid (chain) curve and the dots (squares)
    are results from the perturbation and exact diagonalization
 methods, respectively with the lowest one (two) unperturbed
 subband (subbands) of each hole state considered.}    
  \label{figzw3}
\end{figure} 

We also calculate the spin splitting for holes in 
Si/SiGe QWs with in-plane kinetic energy 
being $k_BT$ ($T=300$~K) along the [100] (solid curve)
and [110] (chain curve) direction in Fig.~\ref{figzw3}(a), 
as done in Ref.~\onlinecite{glavin}. The dotted and dashed curves are taken from
Ref.~\onlinecite{glavin}, corresponding to the spin
splittings along the [100] and [110] directions, respectively. 
It is shown that our
results differ from those in Ref.~\onlinecite{glavin}.\cite{comment} 

We examine our results by further carrying out a calculation of
spin-orbit coupling coefficients/spin splitting with the envelope functions $\Psi_{\lambda\alpha n}$
obtained by the nondegenerate perturbation method as that in
Ref.~\onlinecite{glavin} (also refer to Appendix~A of this paper), 
but with the spin-orbit coupling
coefficients obtained in this work [i.e.,
Eqs.~(\ref{xi}-\ref{theta}) and (\ref{lambda})]. As a 
comparison, we plot the electric field dependence of $\Pi$
[Eq.~(\ref{pi})] in Fig.~\ref{figzw3}(b), where the solid (chain) curve and 
the dots (squares) are results from the perturbation and 
exact diagonalization methods, respectively with the lowest one (two)
unperturbed subband (subbands) of each hole state considered. We find
that when only the lowest unperturbed subband of each hole state is
considered, the perturbation calculation and our exact diagonalization calculation yield almost the identical
results [compare the solid curve
 and the  dots in Fig.~\ref{figzw3}(b)].
 We also find that, as said above,
when more unperturbed subbands of each hole state are accounted, the
nondegenerate perturbation method may cause divergence in the spin-orbit
coupling coefficients/spin splitting (a divergence near $E=500$~kV/cm
in the chain curve is observed). The divergence is caused by 
the degeneracy of the LH subband and the
 SO subband (refer to Appendix~A for details), and disappears in the
exact diagonalization calculation (see squares in the figure).
 Besides, the large discrepancy between the
results with different number of unperturbed subbands included indicates that only considering the lowest subband of each hole state is inadequate for the convergence of calculation. As a result, the exact diagonalization calculation with sufficient unperturbed subbands of each hole state
included is necessary.

\section{Hole spin relaxation}
We perform the fully microscopic KSBE approach\cite{wu-rev} to
study the hole spin relaxation. The KSBEs constructed by
the nonequilibrium Green function method read\cite{wu-rev}
\begin{equation}
  \frac{\partial}{\partial t}\rho_{\bf
    k}(t)=\left.\frac{\partial}{\partial t}\rho_{\bf k}(t)\right|_{\rm coh}
  +\left.\frac{\partial}{\partial t}\rho_{\bf k}(t)\right|_{\rm scat},
\end{equation}
in which $\rho_{\bf k}$  represent the
density matrices of holes.
$\left.\frac{\partial}{\partial t}\rho_{\bf k}(t)\right|_{\rm coh}$ are the coherent terms describing the
coherent spin precessions due to the effective magnetic
fields from the Rashba term and the Hartree-Fock Coulomb interaction.
$\left.\frac{\partial}{\partial t}\rho_{\bf k}(t)\right|_{\rm scat}$
stand for the scattering terms, including the hole-deformation optical/acoustic
phonon,\cite{bufler} hole-impurity
 and hole-hole Coulomb scatterings. Expressions of the
coherent and scattering terms are given in
detail in Ref.~\onlinecite{weng}. What need
to be specified are the matrix elements of the hole-phonon interaction
in the scattering terms. The matrix elements of hole-deformation acoustic phonon scattering
and  hole-deformation optical phonon scattering are
 $|M_{ac,{\bf Q}}|^2=\frac{\hslash
  D_{ac}^2Q}{2dv_s}|I(iq_{z})|^{2}$ and $|M_{op,{\bf Q}}|^2=
\frac{\hslash \Delta_{op}^2}{2d\omega_{op}}|I(iq_{z})|^{2}$, respectively.
 Here ${\bf Q}=({\bf q}, q_z)$ is the phonon momentum.
 The values of mass density $d$, deformation potentials $D_{ac}$ and
 $\Delta_{op}$, the optical phonon energy $\hbar\omega_{op}$ and the sound velocity $v_s$
 in Si and Ge are listed in Table~\ref{table1}. $|I(iq_z)|^2$ is the form factor with
 $I(iq_z)=\int_{-\infty}^{\infty}\Psi_{1l0}^\dagger(z)e^{iq_zz}\Psi_{1l0}(z)dz=\langle\chi_0^{(l1)}(z)|e^{iq_zz}|\chi_0^{(l1)}(z)\rangle+\langle\chi_0^{(l2)}(z)|e^{iq_zz}|\chi_0^{(l2)}(z)\rangle$ for Si/SiGe QWs and
 $I(iq_z)=\int_{-\infty}^{\infty}\Psi_{1h0}^\dagger(z)e^{iq_zz}\Psi_{1h0}(z)dz=\langle\chi_0^{(h)}(z)|e^{iq_zz}|\chi_0^{(h)}(z)\rangle$ for Ge/SiGe QWs. By numerically solving the KSBEs, one can obtain the time evolution 
of density matrices and then the spin relaxation time. In the calculation,
the initial spin polarization of holes is set to be 5~\%.

\subsection{Hole spin relaxation in Si/SiGe QWs}
\begin{figure}[htb]
    {\includegraphics[width=8cm]{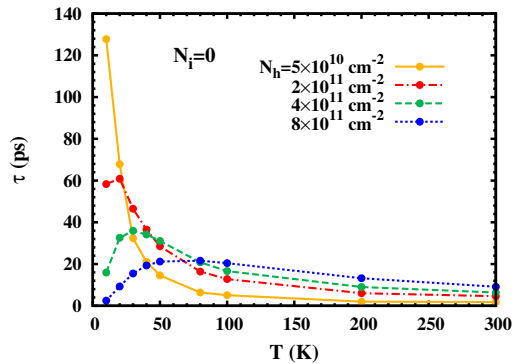}}
    \caption{(Color online) Spin relaxation time $\tau$ against
      temperature $T$ with different hole densities. The impurity
      density $N_i=0$ and the electric field $E=300$~kV/cm.}    
  \label{figzw4}
\end{figure}

\begin{figure}[htb]
    {\includegraphics[width=8cm]{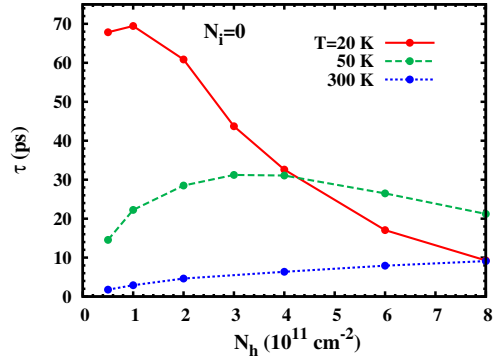}}
    \caption{(Color online) Spin relaxation time $\tau$ against
      hole density $N_h$ with different temperatures. The impurity
      density $N_i=0$ and the electric field $E=300$~kV/cm.}    
  \label{figzw5}
\end{figure}

\begin{figure}[htb]
    {\includegraphics[width=8cm]{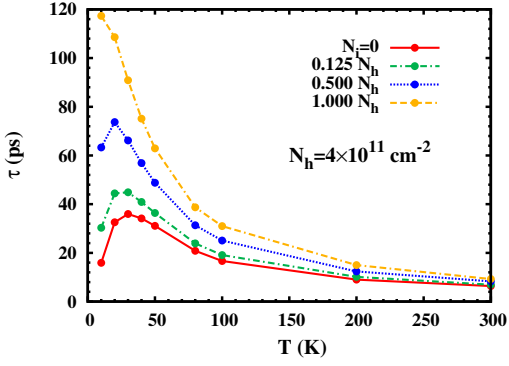}}
    \caption{(Color online) Spin relaxation time $\tau$ against
      temperature $T$ with different impurity densities. The hole
      density $N_h=4\times 10^{11}$~cm$^{-2}$ and the electric field $E=300$~kV/cm.}    
  \label{figzw6}
\end{figure}

\begin{figure}[htb]
    {\includegraphics[width=8cm]{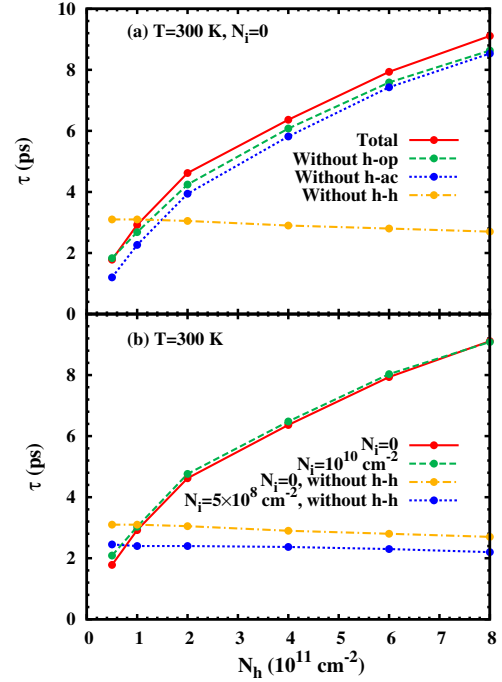}}
    \caption{(Color online) Spin relaxation time $\tau$ against
      hole density $N_h$ with different scatterings
      included.  $T=300$~K and $E=300$~kV/cm. (a) Solid curve: with all
      scatterings [the hole-hole Coulomb (h-h), the hole-optical phonon (h-op) and the hole-acoustic phonon (h-ac)
      scatterings] included; Dashed curve: without the h-op scattering;
Dotted curve: without the h-ac scattering; Chain curve:
  without the h-h scattering. All the four cases are calculated with impurity
 density $N_i=0$. (b) Solid curve: with the h-h, h-op and
h-ac scatterings; Dashed curve: same as the solid curve 
with the additional  hole-impurity scattering
 ($N_i=10^{10}$~cm$^{-2}$) included; Chain curve: with the h-op and
h-ac scatterings; Dotted curve:  same as the chain curve 
with the additional  hole-impurity scattering ($N_i=5\times 10^{8}$~cm$^{-2}$) 
included.}  
  \label{figzw7}
\end{figure}

We first study the spin relaxation of the lowest hole subband in
Si/Si$_{0.7}$Ge$_{0.3}$ QWs. $E=300$~kV/cm unless otherwise specified. 
The LH0 holes have a distribution along the $z$-direction as shown in
Fig.~\ref{figzw1}(a) by the chain curve. The spin-orbit coupling coefficients are 
$\frac{\hbar}{2}\Xi=1.33$~meV~nm, 
$\frac{\hbar}{2}\Pi=-0.83$~meV~nm$^3$ and
$\frac{\hbar}{2}\Theta=-0.39$~meV~nm$^3$, respectively
[Fig.~\ref{figzw2}(b)]. Moreover, due to the quite small material
  parameter $B$ (or $\gamma_2$) in Si, the linear part of
the  Rashba term in Si/SiGe
QWs is relatively more important.  The main results are plotted in
 Figs.~\ref{figzw4}--\ref{figzw7}, showing the spin relaxation with different
 temperatures, carrier/impurity densities and scatterings. 

Figure~\ref{figzw4} shows the temperature dependence of spin relaxation
time with different hole densities. The impurity density is set to be 
zero. A peak, appearing at a temperature
around the Fermi temperature $T_f^h\equiv E_f^h/k_B$ ($E_f^h$ is the
hole Fermi energy. $T_f^h\approx 35$~K with density
$N_h=4\times 10^{11}$~cm$^{-2}$), is observed except when the
  hole density is too low. This kind of
peak has been predicted by Zhou 
{\sl et al.} in high-mobility $n$-doped GaAs QWs
\cite{zhou} and later observed by Ruan {\sl et al.}
experimentally at about $T_f^e/2$ in the temperature dependence of electron spin
relaxation.\cite{ruan} Similar peaks have also
been predicted very recently in the temperature dependence of electron spin
relaxation at a temperature in the range of 
($T_f^e/4$, $T_f^e/2$) in intrinsic bulk GaAs\cite{jiang} 
and at a temperature around the hole Fermi
temperature $T_f^h$ in impurity-free $p$-type GaAs QWs where the hole
density is much higher than the electron density.\cite{yzhou} In fact,
this feature appears in the electron spin relaxation of strong 
scattering system with the DP relaxation mechanism being dominant
when the Coulomb scattering (either the intraband electron-electron or
the interband electron-hole Coulomb
scattering) is the main scattering. 
When the intraband electron-electron Coulomb 
scattering dominates,\cite{zhou,ruan,jiang} the peak appears around the 
crossover from the degenerate to nondegenerate regime of 
electrons. When the interband Coulomb scattering dominates, the peak 
appears around the crossover from the degenerate to nondegenerate regime of
holes in $p$-type systems.\cite{yzhou}

It is known that in the strong scattering system,
strengthening scattering 
can suppress the inhomogeneous broadening and tends to prolong the spin
relaxation time within the DP relaxation mechanism,\cite{wu-rev,zhou,jiang,weng,yzhou,lv1} and that the Coulomb scattering
rate has a $T^{2}$ dependence in the degenerate regime
but a $T^{-1}$ ($T^{-3/2}$) dependence in the nondegenerate
regime in the two (three)-dimensional carrier
  systems.\cite{glazov,giu} Thus with the increase of $T$, the
dominant Coulomb scattering 
tends to cause first an increase and then a
decrease in the electron spin relaxation time. Meanwhile, the increase of
inhomogeneous broadening with $T$ tends to cause a monotonous decrease in
the spin relaxation time and thus  a shift of the
  peak in the $\tau$-$T$ curve towards the lower temperature. The
  magnitude of the latter effect depends on the form of the momentum
  dependence of the DP term. When the DP term mainly depends on
the  momentum linearly (cubically), the latter effect is moderate (strong).
The above scenario also holds in the hole spin relaxation, such as the
 case considered here. The hole system in Si/SiGe QWs is in the strong
scattering limit where the Coulomb scattering dominates (as
discussed later) and the linear part of
the Rashba term is more important, 
thus the peak in the $\tau$-$T$ curve is obvious near
$T_f^h$. However, when the hole density is low enough and thus holes are in the
nondegenerate regime throughout the temperature regime under
consideration, the spin relaxation time decreases monotonously with temperature,
 as shown by the solid curve in Fig.~\ref{figzw4}.

A similar phenomenon is expected to happen in the density dependence 
of spin relaxation time, as the
Coulomb scattering rate has an $N_h^{-1}$ ($N_h^{-2/3}$) 
dependence in the degenerate regime
while an $N_h$ dependence in the nondegenerate
regime in two (three)-dimensional systems.\cite{glazov,giu} This is
exactly the case,\cite{jiang,yzhou} as shown by
Fig.~\ref{figzw5}. Nevertheless, when the temperate is high 
enough and thus holes are in the
nondegenerate regime throughout the density range under
consideration, the spin relaxation time increases monotonously with density, as
shown by the dotted curve in Fig.~\ref{figzw5}.

In Fig.~\ref{figzw6} the spin relaxation time against
temperature with different impurity densities is plotted. It shows
 that adding impurities reduces spin relaxation rate. That is
because the inhomogeneous broadening is suppressed by introducing
hole-impurity scattering in the strong scattering
limit.\cite{wu-rev,zhou,jiang,weng,lv1} With the increase of impurity
density $N_i$, the hole-impurity scattering, which is insensitive to $T$ in low
temperature regime, becomes important. Thus the
peak in $\tau$-$T$ curve due to the Coulomb scattering becomes
less pronounced or even disappears.\cite{zhou,jiang} That is the
reason why the peak is easier to
be observed experimentally in high-mobility samples.\cite{zhou,ruan} It is also noted that
the impurity scattering has marginal effect on spin relaxation
near room temperature, which is understood by recalling that the impurity 
scattering rate decreases with carrier energy.\cite{tomizawa}

To understand the relative importance of different scatterings in spin
relaxation, we calculate the spin relaxation time with different 
scatterings included and show its density dependence in Fig.~\ref{figzw7}. $T$ is
taken to be 300~K. In Fig.~\ref{figzw7}(a) the
impurity density $N_i=0$. The solid curve corresponds to the
case with all the scatterings (the hole-hole Coulomb, hole-optical
phonon and hole-acoustic phonon scatterings) included. 
The dashed, dotted and chain curves
correspond to the cases without the hole-optical phonon,
hole-acoustic phonon and  hole-hole Coulomb scattering, respectively. 
By comparing these four curves,
one finds that: (i) Even with $T=300$~K, the hole-hole Coulomb
scattering plays a much more important role than the hole-phonon
scattering (in addition, similar calculations show that when $T\le 200$~K, the
hole-phonon scattering can be completely ignored); (ii) The acoustic phonon
scattering plays a relatively more efficient role than the optical phonon
scattering [that is because the optical phonon energy
$\hbar\omega_{op}$ is high (63~meV) while the hole
Fermi energy is low due to the large in-plane effective mass (0.27$m_0$)].
In fact, the hole system under consideration is
in the strong scattering limit generally, but falls into the weak scattering
limit when the hole-hole Coulomb scattering is removed artificially. (When
$T=300$~K here, the momentum relaxation time $\tau_p$ due to the 
relatively stronger hole-acoustic phonon scattering is
 about 2.1~ps, while the mean spin precession rate\cite{wu-rev,zhou,lv1}
$\langle\Omega^{(l)}\rangle$ is about 0.52$\sim$0.56~ps$^{-1}$
with the change of hole density. Thus without the hole-hole Coulomb scattering,
$\tau_p\langle\Omega^{(l)}\rangle\gtrsim 1$, indicating the weak
scattering limit.) This feature can be  
justified by comparing two groups of curves in Fig.~\ref{figzw7}(b). One group
with the hole-hole Coulomb scattering (the solid and dashed curves) indicates that
adding a small amount of impurities helps 
increasing $\tau$, corresponding to the strong scattering case, while
the other group without the hole-hole Coulomb scattering (the chain
and dotted curves) shows an inverse effect (i.e., a decrease in $\tau$) with adding quite a small
amount of impurities, which typically happens in the weak
scattering limit.\cite{wu-rev,lv1}

\begin{figure}[htb]
    {\includegraphics[width=8cm]{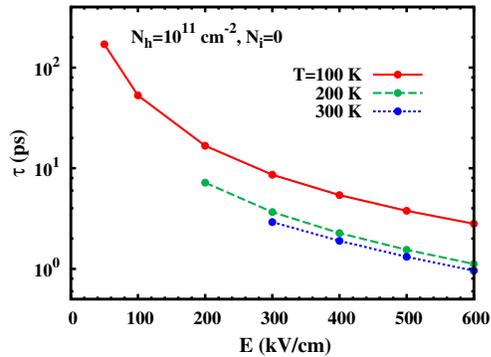}}
    \caption{(Color online) Spin relaxation time $\tau$  against electric field
      $E$ with different temperatures. The hole density is
      $N_h=10^{11}$~cm$^{-2}$ and the impurity density $N_i=0$.}    
  \label{figzw8}
\end{figure} 

\begin{figure}[htb]
    {\includegraphics[width=8cm]{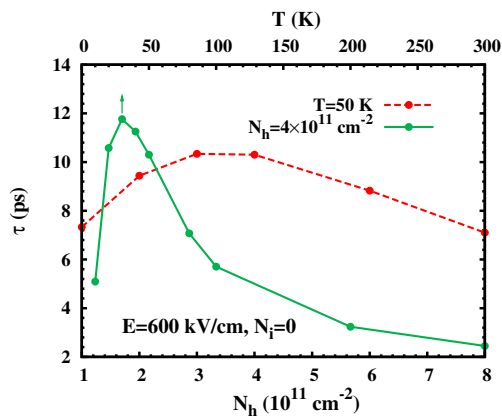}}
    \caption{(Color online) Spin relaxation with electric field
      $E=600$~kV/cm. Dashed curve: spin relaxation time $\tau$ against hole
      density $N_h$ with temperature $T=50$~K; Solid curve: 
spin relaxation time $\tau$ against temperature $T$  with hole
      density $N_h=4\times 10^{11}$~cm$^{-2}$ (Note the
 scale is on the top of the frame). $N_i=0$.}    
  \label{figzw9}
\end{figure}

Finally, we investigate the electric field dependence of spin
relaxation. The hole spin relaxation time $\tau$ against electric field
$E$ with different temperatures is plotted in Fig.~\ref{figzw8}. 
For each temperature, we choose the appropriate range of
electric field to ensure that the effect of second hole subband is irrelevant.
It is shown that with the increase of electric field, the spin relaxation
time is shortened due to the strengthened spin-orbit
coupling. Besides, as a comparison to the features of spin
 relaxation with $E=300$~kV/cm, we 
also present the hole density and temperature dependences of spin
relaxation with $E=600$~kV/cm in Fig.~\ref{figzw9}.
One finds that the peak in hole
density/temperature dependence of the hole spin relaxation time (the
dashed/solid curve in Fig.~\ref{figzw9}) due to the
Coulomb scattering still exists. Moreover, the location of the peak remains
almost the same despite the change of the gate voltage. This indicates
that when the linear part of the Rashba term is important, the trend of variation 
of the hole spin relaxation
time mainly associates with that of the Coulomb scattering
strength around the crossover between the degenerate and
nondegenerate regimes, whereas the increase of the inhomogeneous 
broadening with  increasing  temperature/hole density 
is moderate (even with the increase of the spin-orbit
coupling coefficients by the larger gate voltage). 

\subsection{Spin relaxation in Ge/SiGe QWs}
\begin{figure}[htb]
    {\includegraphics[width=8cm]{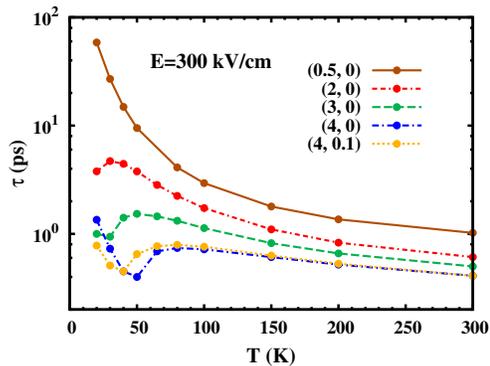}}
    \caption{(Color online) Spin relaxation time $\tau$  against
    temperature $T$ in Ge/SiGe QWs. The electric field $E=300$~kV/cm.
 The curve labeled by $(a, b)$
 corresponds to the case with hole density
 $N_h=a\times 10^{11}$~cm$^{-2}$ and impurity
      density $N_i=b\times 10^{11}$~cm$^{-2}$. The electric field is
      $E=300$~kV/cm. }    
  \label{figzw10}
\end{figure}

The hole spin relaxation in Ge/Si$_{0.3}$Ge$_{0.7}$ QWs is also 
investigated. The HH0 holes have a distribution along the 
$z$-direction as shown in
Fig.~\ref{figzw1}(b) by the chain curve when the electric field is
$E=300$~kV/cm. In Fig.~\ref{figzw10} the spin relaxation time
$\tau$ against temperature $T$ with electric field $E=300$~kV/cm [at
which $\frac{\hbar}{2}\Lambda=-6.06$~meV~nm$^3$, as shown in
Fig.~\ref{figzw2}(b)] is
plotted. It is shown that the hole spin relaxation time in Ge/SiGe QWs
is much shorter than that in
Si/SiGe QWs. That is because the inhomogeneous broadening in Ge/SiGe
QWs is quite strong, as the spin-orbit interaction in
Ge/SiGe QWs is relatively stronger (in consistence with the
heavier Ge element) and the Rashba term depends on
momentum cubically. Apart from the fast spin 
relaxation, a new phenomenon emerges -- when the hole density is
relatively high, not only a peak in $\tau$-$T$
curve is present, but also a valley before the peak is observed.

\begin{figure}[htb]
    {\includegraphics[width=8cm]{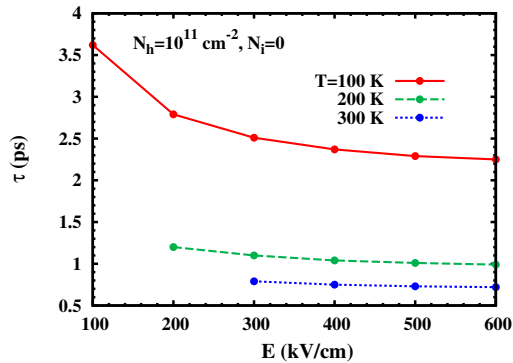}}
    \caption{(Color online) Spin relaxation time $\tau$  against electric field
    $E$ at different temperatures. The hole density is
    $N_h=10^{11}$~cm$^{-2}$ and the impurity density $N_i$ is
      zero.}    
  \label{figzw11}
\end{figure} 

Unlike the case in Si/SiGe QWs where the peak is close to the Fermi
temperature (Fig.~\ref{figzw4}), the peak here is located at a temperature about
$T_f^h/2$ (e.g., $T_f^h\approx 170$~K with
density $N_h=4\times 10^{11}$~cm$^{-2}$). As discussed previously about the
spin relaxation in Si/SiGe QWs, the Coulomb scattering strength first
increases and then decreases with increasing temperature 
(accompanying the crossover from the degenerate to nondegenerate
regime), tending to cause a peak in the $\tau$-$T$ curve around the Fermi
temperature in the  strong scattering limit. However,
 the enhancement of inhomogeneous broadening with $T$ is strong
here as the Rashba term depends on momentum
cubically (unlike the case in Si/SiGe QWs where the linear part of the
Rashba term is important). Thus with both effects accounted, the shift of
the peak towards a lower temperature is expected. 
By comparing the spin relaxation in high temperature regime
of the three curves with $N_i=0$ in Fig.~\ref{figzw10}, 
one finds that the spin relaxation
time $\tau$ decreases with hole density $N_h$ monotonically, 
which is different from the  case in
Si/SiGe QWs (indicated by the dotted curve in Fig~\ref{figzw5}). That is
because the Rashba term here depends on the momentum cubically
and the inhomogeneous broadening increases with density
strongly with an $N_h^3$ dependence.

While the peak is associated with the crossover from the degenerate
to nondegenerate regime, 
the valley in Fig.~\ref{figzw10} is actually related to the crossover from the
weak to strong scattering limit. When the
temperature is low enough and the density is high, the hole
system is highly degenerate (due to the high Fermi energy) and thus the dominant Coulomb
scattering becomes very weak. The  hole system then falls
into the weak scattering limit. Therefore with the increase of
temperature from the very low temperature, the
strengthening of scattering reduces the spin relaxation time first, until the crossover to
the strong scattering system, and then increases the spin relaxation time -- this
leads to a valley in the $\tau$-$T$ curve in the degenerate regime. The dotted curve in
Fig.~\ref{figzw10} stands for the case with the same hole density as
the chain curve but with a small amount of impurities. It is
observed again that introducing a weak impurity scattering in the weak
 (strong) scattering limit leads to a decrease (an increase) of the spin
  relaxation time.\cite{wu-rev,lv1}

The electric field dependence of spin
relaxation is also investigated, with the hole 
spin relaxation time $\tau$ against electric field
$E$ under different temperatures plotted in Fig.~\ref{figzw11}. It
is shown that the spin relaxation
time decreases with electric field slowly in the large electric field regime, corresponding to the
marginal electric field dependence of the spin-orbit coupling strength in the large electric field regime
[refer to the dotted curve in Fig.~\ref{figzw2}(b)].

\section{Conclusion}
In conclusion, we have investigated the hole spin relaxation in [001] strained
asymmetric Si/Si$_{0.7}$Ge$_{0.3}$ (Ge/Si$_{0.3}$Ge$_{0.7}$) QWs with large gate voltage in this
work. We focus on the situations with only the lowest hole subband
being relevant. The effective Hamiltonian of the lowest
hole subband is obtained by the subband L\"owdin perturbation method
in the framework of the six-band Luttinger ${\bf k}\cdot{\bf p}$
model, with sufficient basis functions
included for the convergence of calculation. Due to the biaxial strain, the
lowest subband in Si/SiGe QWs is a light hole-like state, while that in
Ge/SiGe QWs is a heavy hole state.

We apply the fully microscopic KSBE approach to investigate the hole spin
relaxation in Si/SiGe (Ge/SiGe) QWs. By means of
 this approach, all the relevant scatterings,
such as the hole-phonon,  hole-impurity and
the hole-hole Coulomb scatterings can be taken into account
explicitly. It is discovered that the hole-phonon scattering is 
very weak compared to the hole-hole Coulomb scattering, even at high temperatures.
This makes the hole-hole Coulomb scattering to play a  very important
role in spin relaxation.
It leads to a peak of spin relaxation time in
both the temperature and carrier density dependences in Si/SiGe QWs. 
The peak is associated with
the crossover from the degenerate to nondegenerate regime of hole
system, and thus locates around the crossover point. However, the
increase of inhomogeneous broadening with temperature/hole density
tends to lead to a shift of the peak towards a lower temperature/hole
density. The magnitude of the shift depends on the form of the
momentum dependence of the Rashba term. For Si/SiGe (Ge/SiGe) QWs, the
Rashba term mainly (only) depends on momentum linearly (cubically), and
thus the shift of the peak is marginal (obvious). 
In addition, in contrast to the GaAs QWs where the
peak in the temperature dependence of the electron spin relaxation
can only be observed for high mobility samples 
with low carrier density,\cite{zhou}
the peak predicted in Si/SiGe (Ge/SiGe) QWs can be observed 
even at high carrier density, thanks to the weak hole-phonon scattering.
The Coulomb
scattering also leads to a valley at low temperature in the temperature
dependence of hole spin relaxation time in Ge/SiGe QWs with high hole
density, which is related to the crossover from the weak to strong scattering
limit. The hole spin relaxation time can be
effectively influenced by the hole-impurity scattering, tending to
weaken the effect of Coulomb scattering mentioned above with the
increase of impurity density. Apart from the abundant temperature and
hole/impurity density dependences, the spin relaxation time decreases
with the gate voltage, accompanying the increase of spin-orbit
coupling strength. 

The hole spin relaxation time, depending on the
temperature, carrier/impurity density and gate voltage, 
is found to be on the order of 1$\sim$100~ps (0.1$\sim$10~ps) in
Si/SiGe (Ge/SiGe) QWs within the scope of our investigation. 
These time scales are much
shorter than the electron spin relaxation time in Si/SiGe QWs (on the order of
$10^{-7}\sim10^{-5}$~s).\cite{tahan,wil,jantsch} A hole spin
relaxation time on the order of
0.1$\sim$1~ps was theoretically reported in $p$-doped GaAs QWs (with
temperature being 100$\sim$300~K, hole
density being $0.5\sim 4.5\times 10^{11}$~cm$^{-2}$, impurity density
being 0$\sim$1 times hole density and gate voltage induced electric
field being about 100 kV/cm)\cite{lv1} and a hole spin relaxation time of 4~ps was
experimentally observed in $n$-doped GaAs QWs (at 10~K).\cite{damen} Thus, generally, the hole
spin relaxation time in Si/SiGe (Ge/SiGe) QWs is longer than
(comparable with) the hole spin relaxation time in GaAs QWs. It should
be pointed out at last that the strain in Si/SiGe (Ge/SiGe) QWs plays
an important role in spin relaxation, as it shifts the energy 
levels of the light hole states
away from the heavy hole ones. If the strain is removed (e.g.,
in SiO$_2$/Si or SiO$_2$/Ge inversion layer), the coupling 
between the light hole and heavy holea
states are strengthened and the hole spin relaxation should be enhanced.

\begin{acknowledgments}
This work was supported by the National Natural Science Foundation of
China under Grant No.\ 10725417, the National Basic
Research Program of China under Grant No.\ 2006CB922005 and the
Knowledge Innovation Project of Chinese Academy of Sciences. We thank 
  K. W. Kim and B. A. Glavin for valuable discussions.
\end{acknowledgments}

\begin{appendix}
\section{Solution  of the envelope functions $\Psi_{\lambda\alpha
    n}$}
The envelope functions $\Psi_{\lambda\alpha n}(z)$ ($\lambda$=1, 2;
 $\alpha$=$h$, $l$, $s$) satisfy the eigen-equation 
\begin{equation}
H_0\Psi_{\lambda\alpha n}(z)=E_n^{(\alpha)}\Psi_{\lambda\alpha n}(z)
\label{eigen} 
\end{equation}
with $H_0=H_L^{(0)}+H_\epsilon+V(z)I_6$. $\Psi_{\lambda h n}(z)$
has only one component $\chi_n^{(h)}$, satisfying
$[\frac{\hbar^2}{2m^{(h)}_{z}}\frac{d^2}{dz^2}+V(z)]\chi_n^{(h)}(z)=E_n^{(h)}\chi_n^{(h)}(z)$ 
with $m^{(h)}_{z}=m_0/(A-B)$, which can be solved directly. The envelope functions $\Psi_{\lambda \beta
  n}(z)$ ($\beta$ =$l$, $s$) have two components, 
with $\chi_n^{(\beta 1)}$ and $\chi_n^{(\beta 2)}$ satisfying
\begin{eqnarray}
H^{(ls)}_0\left(\begin{array}{c}\chi_n^{(\beta 1)}(z)\\\chi_n^{(\beta 2)}(z)\end{array}\right)&=&E_n^{(\beta)}\left(\begin{array}{c}\chi_n^{(\beta 1)}(z)\\\chi_n^{(\beta 2)}(z)\end{array}\right).
\label{eigenfunction}
\end{eqnarray}
Here
\begin{eqnarray}\nonumber
H^{(ls)}_0&=&\left(\begin{array}{cc}
    (A+B)\frac{\hbar^2}{2m_0}\frac{d^2}{dz^2} &
    -\sqrt{2}B\frac{\hbar^2}{2m_0}\frac{d^2}{dz^2} \\ 
    -\sqrt{2}B\frac{\hbar^2}{2m_0}\frac{d^2}{dz^2} &
    A\frac{\hbar^2}{2m_0}\frac{d^2}{dz^2}-\Delta\end{array}\right)\\
    &&+\left(\begin{array}{cc} E_\epsilon & -\frac{E_\epsilon}{\sqrt{2}} \\ 
    -\frac{E_\epsilon}{\sqrt{2}} & \frac{E_\epsilon}{2}
  \end{array}\right)+V(z)I_2,
\label{hls}
\end{eqnarray}
with the second term representing the contribution from the biaxial
strain.\cite{luttinger1,luttinger2,bir,chao}
$E_\epsilon=-2b(2c_{12}/c_{11}+1)\delta$, where $c_{11}$ and $c_{12}$
are the elastic constants, $b$ is the deformation potential constant 
and $\delta$ is the relative lattice
mismatch in the interface.\cite{chao,schaffler} $E_\epsilon$ is 95.8~meV 
($-115.7$~meV)  for Si/Si$_{0.7}$Ge$_{0.3}$ (Ge/Si$_{0.3}$Ge$_{0.7}$) QWs. 
$\Delta$ is the
SO splitting. The solutions with $\langle\chi_n^{(\beta 1)}(z)|\chi_n^{(\beta 1)}(z)\rangle>\langle\chi_n^{(\beta 2)}(z)|\chi_n^{(\beta 2)}(z)\rangle$ 
correspond to the LH-like states and thus $\beta$ equals to $l$. Otherwise
 the solutions are
deemed as the SO-like states with $\beta=s$. 

An unitary transformation which diagonalizes the strain term in
Eq.~(\ref{hls}) is performed on
Eq.~(\ref{eigenfunction}), leading to $\widetilde{H}^{(ls)}_0$ as
\begin{eqnarray}\nonumber
\widetilde{H}^{(ls)}_0&=&U^{-1}H^{ls}_0U \\\nonumber &=&\left(\begin{array}{cc}
    \frac{\hbar^2}{2m_1}\frac{d^2}{dz^2}+E_1 &
    -\frac{\hbar^2}{2m^\ast}\frac{d^2}{dz^2} \\ 
    -\frac{\hbar^2}{2m^\ast}\frac{d^2}{dz^2} &
    \frac{\hbar^2}{2m_2}\frac{d^2}{dz^2}+E_2 \end{array}\right)\\ &&
+V(z)I_2.
\label{th}
\end{eqnarray}
Here 
\begin{eqnarray}
  U=\left(\begin{array}{cc}
      \frac{1}{\sqrt{N_1}} & \frac{1}{\sqrt{N_2}} \\
      \sqrt{\frac{2}{N_1}}(1-\frac{E_1}{E_\epsilon}) &  
\sqrt{\frac{2}{N_2}}(1-\frac{E_2}{E_\epsilon})  \end{array}\right)
\end{eqnarray}
is the unitary matrix, with $E_1=\frac{1}{2}(\frac{3}{2}E_\epsilon-\Delta+\sqrt{9E_\epsilon^2/4+\Delta
  E_\epsilon+\Delta^2})$, $E_2=\frac{1}{2}(\frac{3}{2}E_\epsilon-\Delta-\sqrt{9E_\epsilon^2/4+\Delta
  E_\epsilon+\Delta^2})$ and
$N_{1,2}=1+2(1-E_{1,2}/E_\epsilon)^2$. $m_1$, $m_2$ and $m^\ast$ in
Eq.~(\ref{th}) are
\begin{eqnarray}\nonumber
m_1&=&m_0\left[A+B\left(\frac{1}{2}+\frac{9E_\epsilon/4+\Delta/2}{\sqrt{9E_\epsilon^2/4+\Delta
  E_\epsilon+\Delta^2}}\right)\right]^{-1},\\\label{effe1}\\\nonumber
m_2&=&m_0\left[A+B\left(\frac{1}{2}-\frac{9E_\epsilon/4+\Delta/2}{\sqrt{9E_\epsilon^2/4+\Delta
        E_\epsilon+\Delta^2}}\right)\right]^{-1},\\ \\
m^\ast&=&m_0\frac{E_\epsilon\sqrt{9/4+\Delta/E_\epsilon+(\Delta/E_\epsilon)^2}}{\sqrt{2}\Delta
  B}.
\label{effe3}
\end{eqnarray}
It is noted that expressions (\ref{effe1}-\ref{effe3}) about the effective masses are valid for
both the Si/SiGe QWs (with $E_\epsilon>0$) and Ge/SiGe QWs (with
$E_\epsilon<0$), while those in Ref.~\onlinecite{glavin} [Eqs.~(5-6)] equal to
Eqs.~(\ref{effe1}-\ref{effe3}) only when $E_\epsilon>0$, i.e., they are 
valid only for Si/SiGe QWs.

$\widetilde{H}^{(ls)}_0$ can be separated into the diagonal 
$\widetilde{H}^{(ls)}_{0D}$ and  off-diagonal $\widetilde{H}^{(ls)}_{0O}$ parts. 
In the nondegenerate perturbation method, $\widetilde{H}^{(ls)}_{0O}$ is treated as the
perturbation term. The eigen-equation of $\widetilde{H}^{(ls)}_{0D}$ can be solved directly,
with the eigen-values $E_{n1}$ and $E_{n2}$ and the corresponding eigen-functions
\begin{eqnarray}
\left(\begin{array}{c} \chi_{n1}(z) \\ 0\end{array}\right) \mbox{and}
\left(\begin{array}{c} 0 \\ \chi_{n2}(z)\end{array}\right)
\label{basis}
\end{eqnarray}
determined by the Schr\"odinger equation
$[\frac{\hbar^2}{2m_\xi}\frac{d^2}{dz^2}+E_\xi+V(z)]\chi_{n\xi}(z)=E_{n\xi}\chi_{n\xi}(z)$
($\xi=1$, 2). In this paper, the term ``unperturbed subbands''
mentioned in the discussion of perturbation 
(exact diagonalization) method actually
means the eigen-states of $\widetilde{H}^{(ls)}_{0D}$, i.e., the
states with energy levels being $E_{n1,2}$ 
and wavefunctions being Eq.~(\ref{basis}).

To the first order of $\widetilde{H}^{(ls)}_{0O}$, the perturbed
eigen-values are $E_n^{(1)}=E_{n1}$ and $E_n^{(2)}=E_{n2}$, and the corresponding
perturbed eigen-functions are 
\begin{eqnarray}
\left(\begin{array}{c} \chi_{n1}(z) \\
    \sum_{n^\prime}\frac{w_{n^\prime2n1}}{E_{n1}-E_{n^\prime 2}}\chi_{n^\prime2}(z)
\end{array}\right) 
\end{eqnarray}
and 
\begin{eqnarray}
\left(\begin{array}{c}
    \sum_{n^\prime}\frac{w_{n^\prime1n2}}{E_{n2}-E_{n^\prime 1}}\chi_{n^\prime1}(z) \\
    \chi_{n2}(z)\end{array}\right),
\end{eqnarray}
respectively. Here $w_{n\xi
  n^\prime\xi^\prime}=-\frac{\hbar^2}{2m^\ast}\int_{-\infty}^{+\infty}dz\chi_{n\xi}\frac{d^2\chi_{n^\prime \xi^\prime}}{dz^2}$ ($\xi$, $\xi^\prime=1$, 2). Finally the eigen-values $E_n^{(\beta)}$ of $H_{0}^{(ls)}$ are $E_n^{(1)}$ and $E_n^{(2)}$, and the corresponding eigen-functions $(\chi_n^{(\beta 1)}(z),\chi_n^{(\beta 2)}(z))^T$ are
\begin{eqnarray}
U\left(\begin{array}{c} \chi_{n1}(z) \\
    \sum_{n^\prime}\frac{w_{n^\prime2n1}}{E_{n1}-E_{n2}}\chi_{n^\prime2}(z)\end{array}\right) 
\label{s1}
\end{eqnarray}
and 
\begin{eqnarray}
U\left(\begin{array}{c}
    \sum_{n^\prime}\frac{w_{n^\prime1n2}}{E_{n2}-E_{n1}}\chi_{n^\prime1}(z) \\
    \chi_{n2}(z)\end{array}\right).
\label{s2}
\end{eqnarray}
 
It is noted that the nondegenerate perturbation method is valid 
when only the two subbands with energy $E_{01}$ and $E_{02}$
  (i.e., the lowest unperturbed subbands corresponding to the LH-like and SO-like hole states respectively) are
accounted. Otherwise, when more subbands are included, the divergence may occur in
the calculation if two energy levels $E_{n1}$
and $E_{n^\prime 2}$ are close to each other, as there are terms proportional to
$\frac{1}{E_{n1}-E_{n^\prime 2}}$ in the envelope functions $\Psi_{\lambda x
  n}(z)$ ($x=l$ and $s$) [refer to
Eqs.~(\ref{s1}--\ref{s2})]. This divergence goes to the spin-orbit coupling coefficients
$\Pi$ and $\Lambda$, and finally the spin splitting. 

For the sake of convergence of the envelope functions and thus the spin-orbit
coupling coefficients and spin splitting, sufficient
subbands have to be considered. Thus,
instead of the nondegenerate perturbation method, we apply the exact
diagonalization method to obtain the eigen-states of $\widetilde{H}_{0}^{(ls)}$, with sufficient basis functions
constructed by the two sets of functions in Eq.~(\ref{basis}) [2 (20) of each set for Si/SiGe (Ge/SiGe) QWs]. 

\end{appendix}


\begin{thebibliography}{0}
\bibitem{note1}The direct energy gap $E_g$ at $\Gamma$ point is much
  larger than the split-off splitting $\Delta$ in Si
  ($E_g=3.4$~eV and $\Delta=0.044$~eV),\cite{low} thus the contribution to the
  spin-splitting of the lowest light hole subband from the conduction
  bands can be neglected. $E_g$ in Ge is relatively smaller
  ($E_g=0.9$~eV and $\Delta=0.3$~eV),\cite{low} but the $\Gamma_{6c}$
  conduction band is found to have no contribution to the
  spin-splitting of the lowest heavy hole subband. Thus the 6$\times$6
  Luttinger model rather than the larger (extended) Kane model\cite{winkler} is
  utilized in the study.
\bibitem{note2}As the electric-field potential is included
  in solving the subband energy spectrum and envelope functions, in the
  basis composed of these subband envelope functions, only the second
  (instead of third\cite{winkler}) order perturbation calculation in L\"owdin partitioning
  is required.
\bibitem{comment} In fact, even 
when the expressions of the spin-orbit
coupling coefficients in Ref.~\onlinecite{glavin} are 
adopted and the nondegenerate
perturbation method with only the lowest unperturbed subband of each
hole state considered is utilized to 
obtain envelope functions $\Psi_{\lambda\alpha n}$, 
the results in Ref.~\onlinecite{glavin} still can not be repeated.
\end{thebibliography}
\end{document}